\documentclass[showpacs,preprintnumbers,superscriptaddress,twocolumn]{revtex4}
\usepackage{amssymb}
\usepackage[centertags]{amsmath}
\usepackage{tipa}
\usepackage{txfonts}
\usepackage{epsfig}
\usepackage{bm}

\begin{document}
\title{Rapidity dependence of  hadron production
in central Au+Au collisions at $\sqrt{s_{NN}}= 200$ GeV}
\author{Jun Song }
\affiliation{Department of Physics, Qufu Normal University,
Shandong 273165, People's Republic of China}

\author{Feng-lan Shao}
\affiliation{Department of Physics, Qufu Normal University,
Shandong 273165, People's Republic of China}

\author{Qu-bing Xie}
\affiliation{Department of Physics, Shandong University,
Shandong 250100, People's Republic of China}

\author{Yun-fei Wang }
\affiliation{Department of Physics, Qufu Normal University,
Shandong 273165, People's Republic of China}

\author{De-ming Wei }
\affiliation{Department of Physics, Qufu Normal University,
Shandong 273165, People's Republic of China}

\begin{abstract}
The rapidity and transverse momentum spectra for identified hadrons
in central Au+Au collisions at $\sqrt{s_{NN}}= 200$ GeV are computed
in a quark combination model. The data of rapidity distributions for
$\pi^{\pm}$, $K^{\pm}$, $p(\overline{p})$ and net protons $(p-\overline{p})$
are well described. We also predict rapidity distributions for $K^{0}_{s}$,
$\Lambda(\overline{\Lambda} )$, $\mathrm{\Xi^{-}}$
($\mathrm{\overline{\Xi}^{\,_+}}$) and
$\mathrm{\Omega^{-}}\!+\mathrm{\overline{\Omega}}^{_+}$. The
multiplicity ratios of charged antihadrons to hadrons as a function
of rapidity are reproduced. The results for ${p}_{T}$  spectra of
$\pi^{\pm}$, $K^{\pm}$, $p(\overline{p})$ and for the $p/\pi$ ratios
in a broader ${p}_{T}$ range agree well with the data. Finally the rapidity
dependence of transverse momentum distributions for hadrons are given.
\end{abstract}


\maketitle

\section{Introduction}

The relativistic heavy ion collider (RHIC) at Brookhaven National
Lab (BNL) provides a unique environment to search for the quark gluon
plasma (QGP) predicted by lattice QCD calculations \cite{Blum:1995},
and to study the properties of this matter at extremely high energy
densities. A huge number of data have been accumulated and used to
extract the information about the original partonic system and its
space-time evolution. A variety of experimental facts from various aspects
imply that the strongly coupled QGP has been probably produced in
central Au+Au collisions at RHIC
\cite{Adams:2005dq,Gyulassy:2004zy,Jacobs:2004qv,Kolb:2003dz,
Braun-Munzinger:2003zd,Rischke:2003mt}. Due to the confinement
effects, one can only detect the hadrons freezed out from the
partonic system  rather than make direct detection of partons
produced in collisions. Therefore, one of the important
prerequisites for exploring the quark gluon plasma is the better
understanding of hadronization mechanism in nucleus-nucleus
collisions.

The quark combination picture is successful in describing many
features of multi-particle production in high energy collisions. The
parton coalescence and recombination models have explained many
highlights at RHIC, such as the high ratio of $p/\pi$ at
intermediate transverse momenta
\cite{Fries:2003,Greco:2003prl,Hwa:2003} and the quark number
scaling of hadron elliptic flows
\cite{Voloshin:2002wa,Molnar:2003ff,Lin:2002rw}. Our quark
combination model has been used to describe the charged particle
pseudorapidity densities \cite{shao:2006}, hadron multiplicity
ratios, $p_{T}$ spectra \cite{Shao:2004cn} and elliptic flows
\cite{Yao:2006fk} at midrapidity.

Recently the STAR collaboration has measured the transverse spectra
for identified baryons and mesons in a wider transverse momentum
range in central Au+Au collisions at $\sqrt{s_{NN}}= 200$ GeV
\cite{Abelev:2006}. The BRAHMS collaboration has given the rapidity
spectra \cite{Bearden:2004y} and the transverse momentum
distributions for identified hadrons at different rapidities
\cite{Arsene:2005,Arsene:2006,Radoslaw:2006}. It provides a good
opportunity to study the hadron production mechanism in a larger
transverse momentum range and at different rapidities. In this
paper, we use our quark combination model to study systematically
the rapidity and transverse momentum distributions for various hadrons
in central Au+Au collisions at $\sqrt{s_{NN}}= 200$ GeV.

The paper is organized as follows. In the next section we give a
brief description of our quark combination model. In section III, the
longitudinal and transverse distributions of partons in central
Au+Au collisions at $\sqrt{s_{NN}}= 200$ GeV are given. In section
IV we compute rapidity distributions of $\pi^{\pm}$, $K^{\pm}$, $p
(\overline{p})$ and the net proton multiplicity $(p-\overline{p})$,
and charged antiparticle-to-particle ratios as a function of
rapidity. We predict the rapidity distributions of $K^{0}_{s}$,
$\Lambda(\overline{\Lambda} )$,\,$\mathrm{\Xi^{-}}$
($\mathrm{\overline{\Xi}^{\,_+}}$) and
$\mathrm{\Omega^{-}}\!+\mathrm{\overline{\Omega}}^{_+}$. In section
V we present our results for ${p}_{T}$ spectra of $\pi^{\pm}$,
$K^{\pm}$ and $p (\overline{p})$. We also give the $p/\pi$ ratios in
a larger ${p}_{T}$ range at $y\sim 0$ and ${p}_{T}$ spectra of
$\pi^{\pm}$ and $p (\overline{p})$ at other rapidities $y\sim1$,
$\eta=2.2$ and $y\thickapprox 3.2$.  Section VI summaries our work.

\section{The quark combination model}
In this section we give a brief description of our quark combination
model. The model was first proposed for high energy $e^+e^-$ and
$pp$ collisions \cite{Xieqb:1988,Liang:1991ya,Wang:1995ch,
Zhao:1995hq,Wang:1996jy,Si:1997ux}. It has also been applied to
the multi-parton systems in high energy $e^+e^-$ annihilations
\cite{Wang:1995gx,Wang:1996pg,Wang:1999xz,Wang:2000bv}. Recently we
have extended the model to ultra-relativistic heavy ion collisions
\cite{Shao:2004cn,Yao:2006fk,shao:2006}.

The average constituent quark number in nucleus-nucleus collisions
can be written as (see e.g. Eq. (4) in Ref.\ \cite{shao:2006})
\begin{equation}
\label{nquark}
\langle{N_q}\rangle=2[(\alpha^{2}+\beta\sqrt{s_{NN}})^{1/2}-\alpha]
\langle{N_{\rm part}}/2\rangle,
\end{equation}
where the two parameters are defined by $\beta={1}/{2\langle
V\rangle}$ and $\alpha=\beta m-\frac{1}{4}$. Here $\langle V\rangle$
is the average inter-quark potential. $m$ is the constituent quark
mass given by $m=(2m_u+\lambda _s m_s)/(2+\lambda _s)$, where $m_u$
and $m_s$ are the constituent masses of light and strange quarks
respectively, and $\lambda _s$ is the strangeness suppression factor.

Our quark combination model describes the hadronization of initially
produced ground state mesons ($36-plets$) and baryons ($56-plets$).
In principle the model can also be applied to the production of
excited states \cite{Wang:1995ch}. These hadrons through combination
of constituent quarks are then allowed to decay into the final state
hadrons. We take into account the decay contributions of all
resonances of $56-plets$ baryons and $36-plets$ mesons, and cover
all available decay channels by using the decay program of PYTHIA
6.1 \cite{Sjostrand}. The main idea is to line up $N_q$ quarks and
anti-quarks in a one-dimensional order in phase space, e.g. in
rapidity, and let them combine into initial hadrons one by one
following a combination rule. See section II of Ref.
\cite{Shao:2004cn} for short description of such a rule.
Of course, we also take into account the near correlation in transverse
momentum by limiting the maximum transverse momentum difference
$\triangle{p_{T}}$ for quarks and antiquarks as they combine into hadrons.
We note
that it is very straightforward to define the combination in one
dimensional phase space, but it is highly complicated to do it in
two or three dimensional phase space \cite{Hofmann:1999jx}.
The
flavor SU(3) symmetry with strangeness suppression in the yields of
initially produced hadrons is fulfilled in the model
\cite{Xieqb:1988,Wang:1995ch}.

\section{longitudinal and transverse distributions of partons}
The BRAHMS Collaboration has measured the ratios of antiparticle-to-particle
and net-baryon rapidity density ($dN_{(B-\overline{B}\,)}/dy$) as a function
of rapidity in central Au+Au collisions at  $\sqrt{s_{NN}}=200$ GeV
\cite{Bearden:2003,bearden:2004}. The results show that the ratios of
$K^{-}/K^{+}$, $\overline{p}/p$ decrease obviously with increasing rapidity
and the $dN_{(B-\overline{B}\,)}/dy$ appears  a valley shape in midrapidity region.
This suggests that the rapidity distribution of net-quarks is different from
that of newborn quarks. In this paper, we describe the momentum distributions
of newborn quarks  by using  a thermal phenomenology with a nonuniform
collective flow. The rapidity distribution of net-quarks is given by
three-sources relativistic diffusion model \cite{Kuiper:2006,wol99}.

\subsection{momentum distributions of newborn quarks}
We start with the  momentum spectrum of quarks radiated by a stationary thermal
source with temperature T
\begin{equation}
\label{sth}
E\frac{d^3n_{th}}{d^3\textbf{p}}=\frac{d^3n_{th}}{dy\hspace{1mm}\mathrm{{p}_{T}}
d\mathrm{{p}_{T}}d\phi}\propto
E\hspace{0.5mm}e^{-E\,/\,T}.
\end{equation}
In the following text we give the spectra in terms of rapidity
$y=\tanh^{-1}(p_{L}/E)$ and transverse momentum $\mathrm{p_{T}}$.

The rapidity distribution of  thermal quarks can be given by
integrating Eq.\,(\ref{sth}) over the transverse components
\cite{Heinz:1993}
\begin{equation}
\begin{split}
\frac{dn_{th}}{dy}&\propto\frac{T^3}{(2\pi)^2}\bigg (\frac{m^2}{T^2}+\frac{m}{T}
\frac{2}{cosh\,y}+\frac{2}{cosh^2y}\bigg)\,\mathrm{exp}\ \bigg(-\frac{m}{T}cosh\,y\bigg),
\end{split}
\label{statth}
\end{equation}
where $m$ is taken to be constituent quark mass, i.e., $m=0.34$ GeV
for light quarks/antiquarks and $m=0.5$ GeV for strange
quarks/antiquarks. Temperature of thermal source is taken to be
$T=170$\ MeV, consistent with the phase transition temperature
($T\sim \,165\,-185$\ MeV ) from lattice QCD calculations.

The momentum distribution of final-state hadrons measured is
 anisotropic in high energy collisions. It rather has
imprinted on  the direction of the colliding nuclei. The
boost-invariant longitudinal expansion scenario
\cite{Bjorken:1983,Heinz:1993} has explained such an anisotropy in
terms of a boost-invariant longitudinal flow of matter with locally
thermalized distribution. We apply the longitudinal flow to the
parton level by adding up the spectra of individual thermal sources
which are distributed in a rapidity region $[y'_{min},y'_{max}]$.
Similar to  nonuniform
longitudinal flow at hadron level \cite{Feng:2000}, we speculate
that the individual thermal sources are nonuniformly distributed in
the rapidity region. The simplest way to describe
this nonuniformity is to introduce a phenomenological  expansion function
 $w(y)$ in the following equation
\begin{equation}
\frac{dn}{dy}(y\,)=N\int\limits_{y'_{min}}^{y'_{max}}\mathrm{d}y'\,w(y')\,
\frac{dn_{th}}{dy}\,(y-y').
\label{expdth}
\end{equation}
\begin{figure}
\epsfig{file=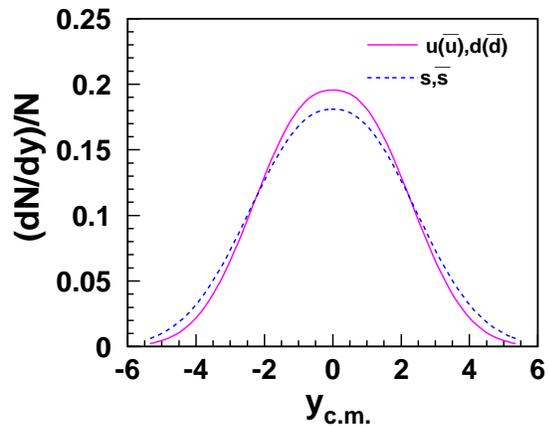,width=\linewidth}
\caption{(Color online) Rapidity spectra of quarks and antiquarks
at hadronization in central Au+Au collisions at $\sqrt{s_{NN}}= 200 $ GeV.}
\label{Fig.1}
\end{figure}
\begin{table}
\caption{ Normalized expansion functions $w(y)$ for light and strange
quarks and antiquarks in central Au+Au collisions at $\sqrt{s_{NN}}= 200 $ GeV.}
\tabcolsep0.15in \arrayrulewidth0.5pt
\renewcommand{\arraystretch}{2.5}
\begin{tabular}[t]{c c}\hline \hline
  $u$,\,$\overline{u}$\ and $d$,\,$\overline{d}$
 &$w(y)=0.2\,\mathrm {exp}\,\Big(-|y|^{2.62}/{14.85}\Big)$ \\ \hline
   s\,, $\overline{s}$  &
 $w(y)=0.184\,\mathrm {exp}\,\Big(-|y|^{2.4}/{14.85}\Big)$\\ \hline \hline
\end{tabular}
\label{wfunc200}
\end{table}
The transverse momentum spectrum of quark is not affected by this
operation \cite{Heinz:1993}. The integrity interval is chosen to be $y'_{max}=
-y'_{min}=y_{beam}$. $N$ is the normalization constant. In the
quark combination model, excluding the influence of
resonance decays from final-state $\pi^{+}$ and $K^{+}$ rapidity
distribution, we get the initially produced $\pi^{+}$ and $K^{+}$
rapidity spectra. We further inversely extract the
expansion functions $w(y)$ for light and strange quarks and
antiquarks, and  obtain rapidity distributions of light and strange
quarks and antiquarks. The results are shown in
Table\,\ref{wfunc200} and Fig.\,\ref{Fig.1} respectively.

The transverse momentum spectrum of  thermal quarks can be
got by integrating Eq.\,(\ref{sth}) over rapidity using the modified
Bessel function $K_{1}$ \cite{Heinz:1993}
\begin{equation}
\label{th-pt}
\frac{dn_{th}}{{2\pi\hspace{1mm}\mathrm{{p}_{T}}
 d\mathrm{{p}_{T}}}}\propto m_{T}K_{1}
\bigg(\frac{m_{T}}{T}\bigg),
\end{equation}
which behaves asymptotically like a decreasing exponential
$\mathrm{exp}\,(-m_{T}/{T})$, here transverse mass $m_{T}=\sqrt{p^2_{T}+m^2}$.

Taken into account transverse flow of thermal source,
the partons transversely boost by a flow velocity profile $\beta_{r}(r)$ as
 a function of transverse radial positions $r$. $\beta_{r}(r)$ is parameterized
by the surface velocity $\beta_{s}$:
$\beta_{r}(r)=\beta_{s}\,(r/R_{max})^{n}$. The transverse
momentum spectrum of expansion thermal source can be described as
 \cite{Heinz:1993}
\begin{equation}
\label{thermpt}
\frac{dn_{th}}{{2\pi\hspace{1mm}\mathrm{{p}_{T}}
 d\mathrm{{p}_{T}}}}=N\int_{0}^{R_{max}}r\,dr\, m_{T}\,I_{0}\bigg(
\frac{p_{T}\,sinh\,\rho}{T}\bigg)K_{1}\bigg(
\frac{m_{T}\,cosh\,\rho}{T}\bigg),
\end{equation}
where $I_{0}$ is modified Bessel function, and $\rho=tanh^{-1}\beta_{r}$ is
transverse rapidity. $N$ is the normalization constant. The value of $R_{max}$
is taken to be $R_{max}=13\,\mathrm{fm}$ \cite{ref:13fm}.
Values of  parameters $n,\,\beta_{s}$ for light and strange quarks and antiquarks
are extracted from the transverse momentum distributions of $\pi^{0}$ and $K^{0}_{s}$
at midrapidity in our quark combination model.

Partons at high transverse momenta are mainly from the minijets
 created in initial hard collisions among nucleons. The transverse momentum
 distribution of minijet partons in the midrapidity region can be parameterized as
\cite{Fries:2003,Greco:2003prc}
\begin{equation}
\label{minjet-pt}
\frac{dn_{jet}}{{2\pi\hspace{1mm}\mathrm{{p}_{T}}
 d\mathrm{{p}_{T}}}}=
 A\,\bigg(\frac{B}{B+\mathrm {p _{T}}}\bigg)^{\,C}.
\end{equation}

\begin{figure}
\epsfig{file=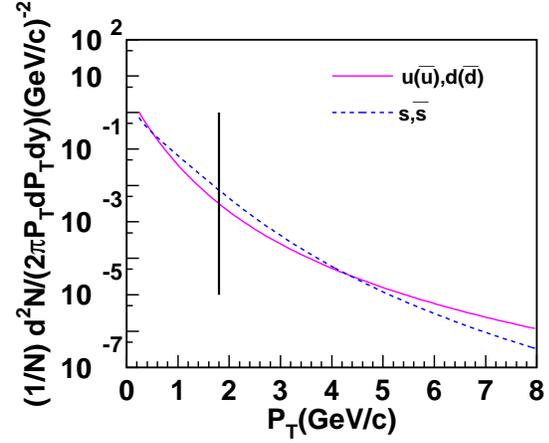,width=\linewidth}
\caption{(Color online) Transverse momentum spectra of quarks and antiquarks
at midrapidity in  central Au+Au collisions at $\sqrt{s_{NN}}= 200$ GeV.
Minijet quarks have transverse momenta $\mathrm {p _{T}}>1.8$ GeV,
while thermal quarks have transverse momenta $\mathrm {p _{T}}<1.8$ GeV.}
\label{qpt}
\end{figure}

 The transition point ($\mathrm {p _{0}}$) from thermal
partons to minijet partons  in $\mathrm {p _{T}}$ spectra of quarks
and antiquarks is determined by the cross point of the two functions in  Eq.\,(\ref{thermpt}) and
Eq.\,(\ref{minjet-pt}). Thermal partons dominate the  transverse momentum below
$\mathrm {p _{0}}$ while minijet partons dominate the transverse momentum greater than
$\mathrm {p _{0}}$. We extract the $\mathrm {p _{T}}$
spectra of quarks and antiquarks from measured $\pi^{0}$ and $K^{0}_{s}$
spectra, then transition point $\mathrm {p _{0}}=1.8$ GeV/c is taken.
For thermal quarks, We take $n=0.5$ for light and strange quarks and antiquarks,
$\beta_{s}=0.45 c,0.56 c$ for light and strange quarks/antiquarks respectively.
Values of parameters A, B and C for minijet partons are given in Table.\,\ref{xif200}.
The scattering of minijet partons with thermal partons may lead to some change
of the spectra of minijet partons and the smooth spectra around $\mathrm {p _{0}}$.
Normalized transverse momentum distributions for light and strange quarks and
antiquarks at midrapidity in central Au+Au collisions at $\sqrt{s_{NN}}= 200$
GeV are in  Fig.\,\ref{qpt}.

\begin{table}
\caption{Parameters for minijet parton distributions given in Eq.(\ref{minjet-pt})
at midrapidity in central Au+Au collisions at $\sqrt{s_{NN}}= 200 $ GeV.}
\tabcolsep0.15in
\setlength{\arrayrulewidth}{.5pt}
\renewcommand{\arraystretch}{1.5}
\begin{tabular}[t]{c c c c}\hline \hline
&$A\,[1/\text{GeV}^2]$&$B\,[1/\text{GeV}]$&$C$ \\ \hline
 $u$,\,$\overline{u}$\ and $d$,\,$\overline{d}$
&$2.32$&1.5&7.25\\ \hline
 s\,, $\overline{s}$&$4.51$&1.5&9.5\\ \hline \hline
\end{tabular}
\label{xif200}
\end{table}

\subsection{rapidity distribution of net-quarks}
 In this subsection, we use the three-sources relativistic diffusion model
(RDM) \cite{Kuiper:2006,wol99} to describe the rapidity distribution of
net-quarks. The time evolution of distribution functions is given by a
Fokker-Planck equation (FPE) in rapidity space
\cite{wol99,wol03,alb00,ryb02,biy02}
\begin{equation}
\frac{\partial}{\partial t}[ R(y,t)]^{\mu}=-\frac{\partial}
{\partial y}\Bigl[J(y)[R(y,t)]^{\mu}\Bigr]+D_{y}
\frac{\partial^2}{\partial y^2}[R(y,t)]^{\nu},
\label{fpenl}
\end{equation}\\
where $J(y)$ is drift function and $D_{y}$ is rapidity diffusion coefficient.

Here, we use $q=\nu=1$ corresponding to the standard FPE, and a linear drift
function \cite{wol99}
\begin{equation}
J(y)=(y_{eq}- y)/\tau_{y},
\label{dri}
\end{equation}
where $\tau_{y}$ is rapidity relaxation time,  $y_{eq}$ is rapidity
equilibrium value and taken to be $y_{eq}=0$ for symmetric systems.

For the linear case above, exact solutions of standard FPE  originating
from $R_{1,2}(y,t=0)=\delta(y\mp y_{beam})$ can be obtained \cite{wol99b}
\begin{equation}
R_{1,2}(y,t)=\frac{1}{\sqrt{2\pi\,\sigma_{1,2}^{2}(t)}}\,
\mathrm{exp}\,\Bigl[-\frac{\big(y-\langle y_{1,2}(t)\rangle\big)^{2}}{2\,
\sigma_{1,2}^{2}(t)}\Bigr],
\label{fpesol}
\end{equation}
where
\begin{equation}
\begin{split}
\langle y_{1,2}(t)\rangle\,= \pm y_{beam}\exp{(-t/\tau_{y})}\\
\sigma_{1,2}^{2}(t)=D_{y}\tau_{y}\,[1-\exp(-2t/\tau_{y})].
\end{split}
\end{equation}

The available net-proton data at AGS ( Au+Au collisions at
$\sqrt{s_{NN}}= 5$ GeV) and SPS (Pb+Pb collisions at $\sqrt{s_{NN}}=17$ GeV)
are reproduced well with a superposition of two sources $R_{1,2}(y,t)$ in RDM.
But it has been proposed in Ref. \cite{wol03} that an expanding midrapidity
source $R_{3}(y)$ emerges at RHIC. With this conjecture, the rapidity
distribution of net-quarks in  Au+Au  collisions at $\sqrt{s_{NN}}= 200 $
GeV can be written  as
\cite{wol99,Kuiper:2006,wol03}
\begin{equation}
\frac{dN(y,t=\tau_{int})}{dy}=N_{1}R_{1}(y,\tau_{int})+N_{2}R_{2}(y,\tau_{int})
+N_{mid}R_{3}(y)
\label{dynetq}
\end{equation}\\
 with the total number of net-quarks $N_{1}+N_{2}+N_{mid}$ equal to
 $3N_{part}$. Here $\tau_{int}$ is interaction time. The first and
second items give the prediction of rapidity distribution for net-quarks
at large rapidity, while the last $R_{3}(y)$  describes the rapidity distribution
of net-quarks in the midrapidity region.

There are only two adjustable parameters ${\tau_{int}}/{\tau_y}$ and
$D_{y}{\tau_y}$ in $R_{1,2}(y,\tau_{int})$. ${\tau_{int}}/{\tau_y}$
represents the diffusive time evolution of  net-quarks; $D_{y}{\tau_y}$
accounts for the widening of the rapidity distributions due to interactions
among partons and partonic creations. In the present paper, $D_{y}{\tau_y}$
has included the enhancement factor $g(\sqrt{s})$  caused by memory and
collective effects \cite{wol99,wols99,wol02}. In central Au+Au collisions
at $\sqrt{s_{NN}}= 200 $ GeV, values for parameters are taken to be
${\tau_{int}}/{\tau_y}=0.263$ and $D_{y}{\tau_y}=0.9$ respectively.

The stationary solution of Eq.\ref{fpenl} is approximate to Gaussian
distribution \cite{wol99b}. It is  regarded as $R_{3}(y)$ to describes
the midrapidity valley for net-baryon rapidity distribution in
Ref.\cite{wol99,wol03,Kuiper:2006}. Instead, for a better description
of rapidity distribution of net-quarks, we obtain $R_{3}(y)$ by fitting
the data of net-baryon in our model. The normalized polynomial fit for
$R_{3}(y)$ in a rapidity region $-y_{beam}<y<y_{beam}$ is
\begin{equation}
\begin{split}
&R_{3}(y)=0.0715376 -0.0189521y^{2} - 0.004336y^{4}\\
&+ 0.0000424407y^{6}+0.01001|y|+0.0194476 |y|^{3}.
\end{split}
\end{equation}
\begin{figure}[h]
\epsfig{file=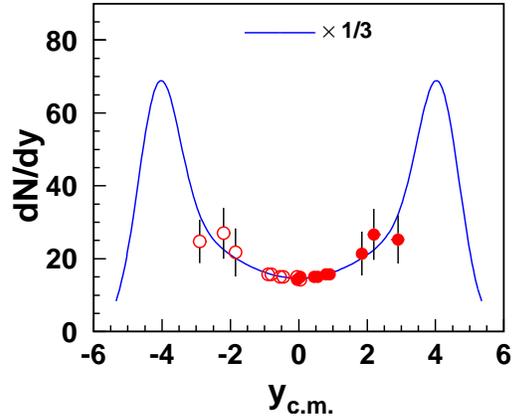,width=\linewidth}
\caption{(Color online) Rapidity distribution of net-quarks in central
Au+Au  collisions at $\sqrt{s_{NN}}= 200 $ GeV. The spectrum is multiplied
by $1/3$ to compare with the net-baryon data from  BRAHMS
Collaboration \cite{bearden:2004}.}
\label{netq200}
\end{figure}

The rapidity distribution of net-quarks in central Au+Au collisions
at $\sqrt{s_{NN}}= 200 $ GeV is shown in Fig.\,\ref{netq200}.
Values of $N_{1},N_{2},N_{mid}$ are taken to be $N_{1}=N_{2}=214$ and
 $N_{mid}=607$ for central Au+Au collisions at $\sqrt{s_{NN}}= 200 $ GeV.

\section{rapidity distributions of identified hadrons}
In this section, we use the quark combination model to compute the
rapidity distributions of identified hadrons in central Au+Au
collisions at $\sqrt{s_{NN}}= 200 $ GeV. The total number of
constituent quark is given in Eq.\,(\ref{nquark}). The values of
parameters $\lambda _s$ and $\beta$ in Eq.\,(\ref{nquark}) are
taken to be $\lambda _s=0.55$, $\beta=0.36\, \texttt{GeV}^{-1}$ respectively
\cite{shao:2006,Shao:2004cn}. The rapidity distributions of newborn
quarks and net-quarks are given respectively in Eq.\,(\ref{expdth})
and Eq.\,(\ref{dynetq}). The number of net-quarks is $3N_{part}$ and
all the net-quarks are allowed to take part in combination.
Minijet partons are indispensable to hadron productions at large transverse
momenta. But from the transverse momentum spectra of partons at midrapidity in Fig.\,\ref{qpt},
we can see that the amount of minijet partons is about two order of magnitude
smaller than that of thermal partons. The contribution from  minijet
partons to the yields and rapidity distributions of hadrons is negligible.
Therefore, we don't distinguish between minijet and thermal partons in rapidity.
With this
input, we can calculate the rapidity distributions of various hadrons.
\begin{figure}[h]
\epsfig{file=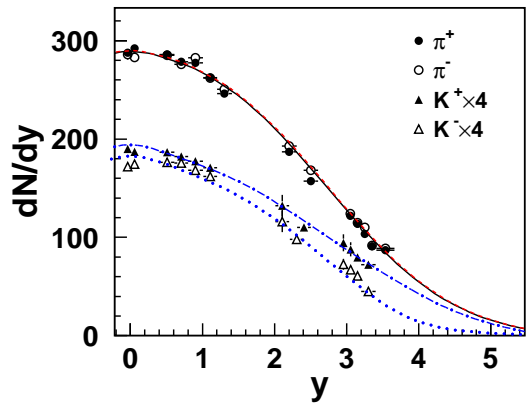,width=\linewidth}
\caption{(Color online) Pions and kaons rapidity densities as a function of
rapidity in central Au+Au collisions at $\sqrt{s_{NN}}= 200 $ GeV. The kaons
yields were multiplied by 4 for clarity. The solid, dashed, dotted-dashed,
dotted lines are our results for $\pi^{+}$, $\pi^{-}$, $K^{+}$ and $K^{-}$
respectively. The experimental data are given by  BRAHMS Collaboration
\cite{Bearden:2004y}.}
\label{Fig.4}
\end{figure}

In Fig.\ref{Fig.4}, we  show the rapidity distributions of charged
pions and kaons in central Au+Au collisions at $\sqrt{s_{NN}}= 200 $
GeV. The pion yields are collected excluding the contribution of
hyperon $(\Lambda)$ and neutral kaon $K^{0}_{s}$ decays.
In quark combination picture, the existence of net-quarks will lead to
the excess of direct produced $K^{+}$ over $K^{-}$  yields, and decay
contributions from other particles $(K^{*0},\phi,\Omega)$ hardly make
change to this excess. Therefore the  excess of $K^{+}$ over $K^{-}$
yields appears in the full rapidity range. While the directly produced
$\pi^{+}$ and $\pi^{-}$ yields can not reflect the influence of net
quarks. Even if the decay contributions from other hadrons are considered,
the difference between yields of $\pi^{+}$ and $\pi^{-}$  is also very small
compared with the large yields of pions. Thus the rapidity densities of
 $\pi^{+}$ and $\pi^{-}$ are  nearly equal within the  entire rapidity region.
One can see that our model  well describes the rapidity densities ${dN}/{dy}$
of pions and kaons in the rapidity range covered for central Au+Au
collisions at $\sqrt{s_{NN}}= 200 $ GeV.

\begin{figure}[h]
\epsfig{file=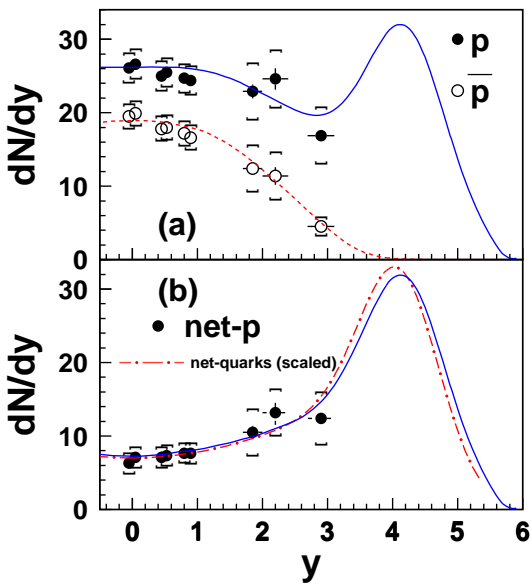,width=\linewidth}
\caption{(Color online) Proton, antiproton and net-proton rapidity densities
${dN}/{dy}$ as a function of rapidity in central Au+Au collisions at
$\sqrt{s_{NN}}= 200 $ GeV. The solid and dashed lines in (a) are our results
for proton, antiproton respectively, the solid line in (b) is our result
for net-proton. The spectrum of net-quarks scaled is compared with that
of net-proton. The errors shown with caps include both statistical and
systematic. The experimental data are given by BRAHMS Collaboration
\cite{Bearden:2004y}.}
\label{Fig.5}
\end{figure}

The rapidity distribution of  net-proton can reflect the energy loss of
colliding nuclei and the degree of  nuclear stopping. We calculate the
rapidity distributions of proton, antiproton, and net-proton in central
Au+Au collisions at $\sqrt{s_{NN}}= 200 $ GeV. The results are shown in
Fig.\ref{Fig.5}.  No weak decay correction has been applied. The existence
of net quarks  also leads to the excess of directly produced proton over
antiproton yields. Taken into account the decay contributions from hyperons,
this excess will further increase. Our quark combination model can also
successfully explain the data of proton, antiproton, and net-proton
in the rapidity region $0<y<3$.

The process of quarks combination is a conversion process of net-quarks
to net-baryons  simultaneously. In Fig.\ref{Fig.5}(b), the computed rapidity
distribution of net-proton nearly agrees with that of
net-quarks except a small change of peak position around $y\thickapprox4$.
This is mainly because of the mass effects in the processes of quarks combination
and resonances decays (see Ref.\,\cite{shao:2006} for explanation of this mass
effects). Other baryons also have the similar character. On the other
hand, the rapidity distributions of hadrons are only slightly affected by
rescattering in late stage of collisions \cite{Boggild:2006}. Therefore,
if quark combination mechanism is also applicable to large rapidity region,
the measured net-baryon ($B-\overline{B}$) rapidity distribution will reflects
quite truly the rapidity distribution of net-quarks just before hadronization. This
provides a very beneficial condition to study the energy loss of colliding
nuclei and the degree of nuclear stopping in nucleus-nucleus collisions at RHIC.

\begin{figure}
\epsfig{file=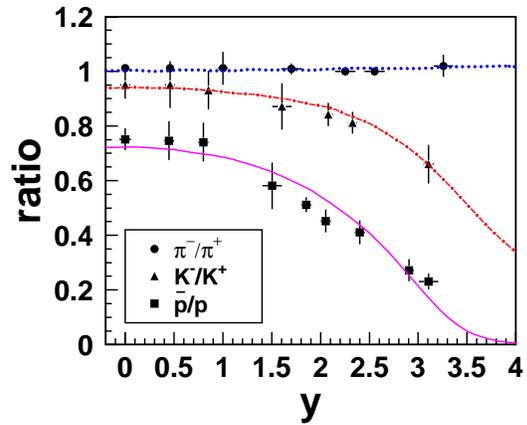,width=\linewidth}
\caption{(Color online) Antiparticle-to-particle ratios as a function of
rapidity in central Au+Au collisions at $\sqrt{s_{NN}}= 200 $ GeV. The
dotted, dotted-dashed and solid lines are our results for $\pi^{-}/\pi^{+}$,
$K^{-}/K^{+}$, $\overline{p}/p$ respectively. The experimental data are from
BRAHMS Collaboration
\cite{Bearden:2003}. }
\label{Fig.6}
\end{figure}

Rapidity dependence of the ratios for antiparticles to particles are significant
indicators of the dynamics of  high energy nucleus-nucleus collisions
\cite{Herrmann:1999,Satz:2000}. Since we have computed the rapidity spectra
of charged pions, kaons, proton and antiproton, it is convenient to compute
the ratios of yields of antiparticles to particles varied with rapidity.
In Fig.\ref{Fig.6}, we show the $\pi^{-}/\pi^{+},\, K^{-}/K^{+}$ and
$\overline{p}/p$ ratios as a function of rapidity in central Au+Au collisions
at $\sqrt{s_{NN}}= 200 $ GeV. No weak decay correction has been applied. The
ratio of $\pi^{-}/\pi^{+}$ is consistent with unity over the entire rapidity
range, while the ratios of $ K^{-}/K^{+}$ and $\overline{p}/p$ decrease with
increasing rapidity due to the influence of net quarks. Our results agree well
with the data.

\begin{figure}
\epsfig{file=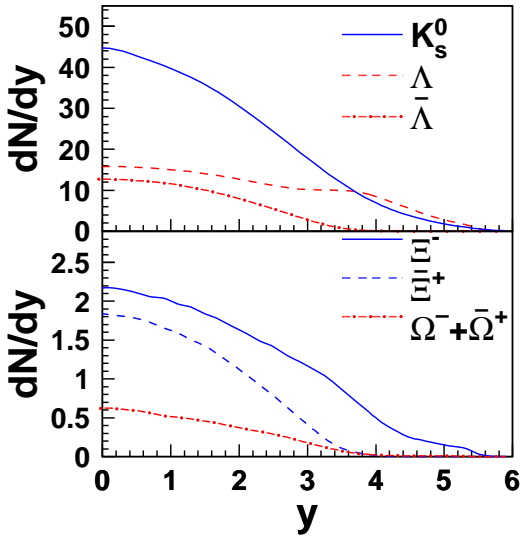,width=\linewidth}
\caption{(Color online) Rapidity distributions for neutral kaon $K^{0}_{s}$ and
hyperons $\Lambda(\overline{\Lambda} )$,\,$\mathrm{\Xi^{-}}$ \!
($\mathrm{\overline{\Xi}^{\,_+}}$),\,$\mathrm{\Omega^{-}}\!+\mathrm{\overline
 {\Omega}}^{_+}$ in central Au+Au collisions at $\sqrt{s_{NN}}= 200 $ GeV.}
\label{ystrange}
\end{figure}

From the above results, one can see that our model can well describe the
rapidity distributions of primary charged hadrons. Among hadrons combined
by light and strange quarks and antiquarks, the productions of hyperons
$\mathrm{\Xi}$ and $\mathrm{\Omega}$ are less influenced by the decays of
other hadrons, thus their rapidity distributions can reflect more directly
the hadronization mechanism. We predict the rapidity spectra of neutral kaon
$K^{0}_{s}$ and hyperons $\Lambda(\overline{\Lambda} )$,\,$\mathrm{\Xi^{-}}$ \!
($\mathrm{\overline{\Xi}^{\,_+}}$),\,$\mathrm{\Omega^{-}}\!+
\mathrm{\overline{\Omega}}^{_+}$ in central Au+Au collisions at $\sqrt{s_{NN}}= 200 $
GeV. The results are shown in Fig.\ref{ystrange}. We also compute the rapidity
densities of these hadrons in midrapidity region and compare the results with the
experimental data. The data and calculation results are in Table \ref{midrapidity}.

\begin{table}
\caption{Rapidity densities ${dN}/{dy}$ for neutral kaon $K^{0}_{s}$ and hyperons
$\Lambda(\overline{\Lambda} )$,\,$\mathrm{\Xi^{-}}$ \! ($\mathrm{\overline
{\Xi}^{\,_+}}$), $\mathrm{\Omega^{-}}\!+\mathrm{\overline{\Omega}}^{_+}$
at midrapidity in central Au+Au collisions at $\sqrt{s_{NN}}= 200 $ GeV. The
value of $K_{s}^{0}$ is the fit result according to its transverse momentum spectrum
given by STAR Collaboration \cite{Admas2006b}, other data are from  STAR Collaboration
\cite{Admas2006a}. }
\tabcolsep0.15in
\setlength{\arrayrulewidth}{.5pt}
\renewcommand{\arraystretch}{1.5}
\begin{tabular}[t]{c c c}\hline \hline
 &data& our model \\ \hline
$K^{0}_{s}$ & $45.28^{\,*}$ & 44.1 \\ \hline
$\Lambda$ &$16.7\pm0.2\pm1.1$ &15.9 \\ \hline
$\overline{\Lambda}$ &$12.7\pm0.2\pm0.9$&12.5 \\  \hline
$\mathrm{\Xi^{-}}$& $2.17\pm0.06\pm0.19$ &2.13 \\ \hline
$\mathrm{\overline{\Xi}^{\,_+}}$ & $1.83\pm0.05\pm0.2$ &1.76 \\ \hline
$\mathrm{\Omega^{-}}\!+\mathrm{\overline{\Omega}}^{_+}$ &$0.53\pm0.04\pm0.04$ & 0.59
\\ \hline \hline
\end{tabular}
\label{midrapidity}
\end{table}
\section{rapidity dependence of transverse momentum spectra for identified hadrons}
In this section, we calculate ${p}_{T}$  spectra of pions, kaons, protons and
the $p/\pi$ ratios in a larger ${p}_{T}$ range at $y\sim0$ in central Au+Au
collisions at $\sqrt{s_{NN}}= 200 $ GeV. We further compute the transverse
momentum distributions of identified hadrons at other rapidities $y\sim1$,
$\eta=2.2$ and $y\thickapprox3.2$.

The transverse momentum distributions of newborn quarks and antiquarks  have
been given in section III. Based on the experimental fact that $\pi^{-}/\pi^{+}$
and  $\overline{p}/p$ ratios at midrapidity are  almost constants within a broad
transverse momentum region \cite{Abelev:2006}, the transverse momentum distribution
of net-quarks is taken the same as that of newborn light quarks at midrapidity.
We use $\triangle{p_{T}}=0.66$ GeV for mesons and $\triangle{p_{T}}=1.0$ GeV
for baryons respectively. With this input, we can compute the transverse momentum
distributions of various hadrons.
\subsection{Transverse momentum distributions of identified hadrons in a wider  $\mathrm {p_{T}}$
 range at $y\sim0$}
Identified baryon and meson distributions at large transverse
momenta are available in Au+Au collisions at $\sqrt{s_{NN}}= 200 $
GeV \cite{Abelev:2006}. It allows one to explore the particle
production mechanism in a larger transverse momentum range. In this
subsection, we compute the transverse spectra of identified hadrons
at large $\mathrm {p_{T}}$ and study the applicable extent of
combination mechanism in a larger transverse momentum range.
 \begin{figure}[h]
\epsfig{file=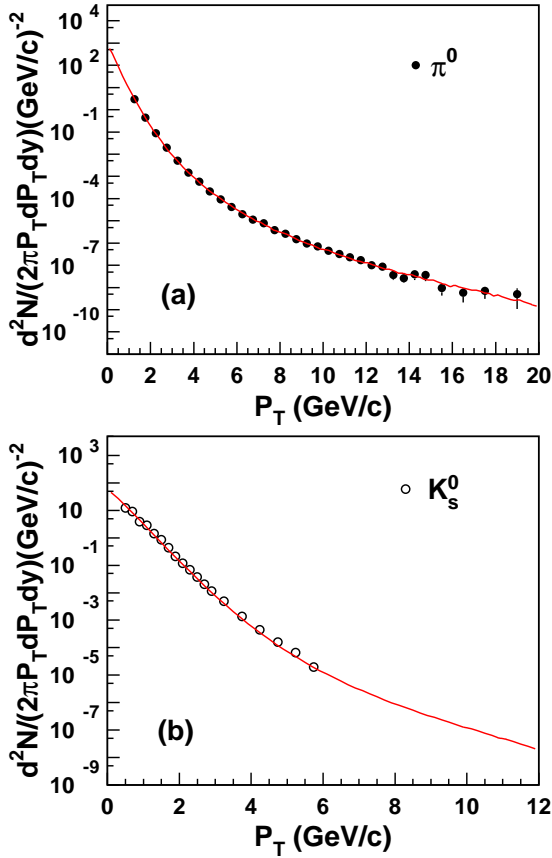,width=\linewidth} \caption{(Color online)
Transverse momentum spectra of $\pi^{0}$ (a), $K^{0}_{s}$ (b) at
midrapidity in central Au+Au  collisions at $\sqrt{s_{NN}}= 200 $
GeV. The solid lines are our results. The data of $\pi^{0}$ are
given by PHENIX Collaboration \cite{Tadaaki:2005}, and data of
$K^{0}_{s}$ are from STAR Collaboration \cite{Admas2006b}.}
  \label{pi-ks0pt}
\end{figure}

\begin{figure*}[!tbp]
\epsfig{file=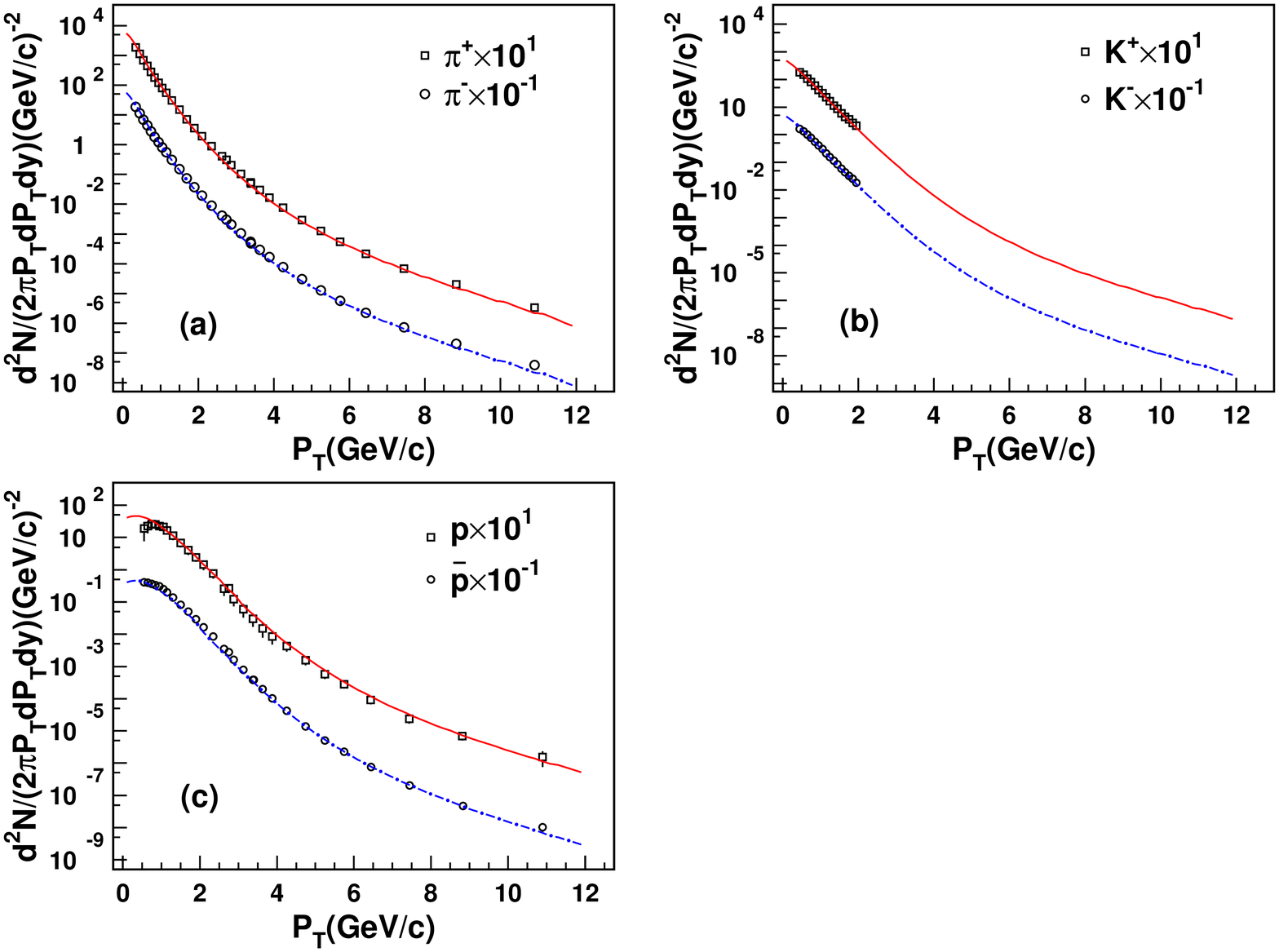,width=\linewidth} \caption{(Color online)
Transverse momentum spectra of pions (a), kaons (b), proton and
antiproton (c) at midrapidity in central Au+Au collisions at
$\sqrt{s_{NN}}= 200 $ GeV. The solid lines are our results for
$\pi^{+}$, $K^{+}$ and $p$, and the dotted-dashed lines are for
$\pi^{-}$, $K^{-}$ and $\overline{p}$ respectively. The data of
pions, proton and antiproton are from
 STAR Collaboration \cite{Abelev:2006}, and kaons are from BRAHMS
Collaboration \cite{Arsene:2005}.} \label{pionpt}
\end{figure*}

In Fig.\ref{pi-ks0pt}, we show transverse momentum spectra of $\pi^{0}$ and
$K^{0}_{s}$ at midrapidity in central Au+Au collisions at $\sqrt{s_{NN}}= 200 $
 GeV, which are used to determine the values of parameters for parton
 $\mathrm {p_{T}}$ spectra. No decay corrections are applied to data.
 The transverse momentum spectra of  pions,
kaons, proton and antiproton  are calculated at midrapidity in central Au+Au
collisions at $\sqrt{s_{NN}}= 200 $ GeV. The results are in Fig.\ref{pionpt}.
The pions spectra are corrected to
remove the feed-down contributions from $K_{s}^{0}$ and $\Lambda$. The feed-down
contributions from $\Lambda$ and $\Sigma^{+}$ are subtracted from the proton
spectrum. Our quark combination  model can
also well explain the data in a large $p_{T}$ range.

\begin{figure}[!tbp]
\epsfig{file=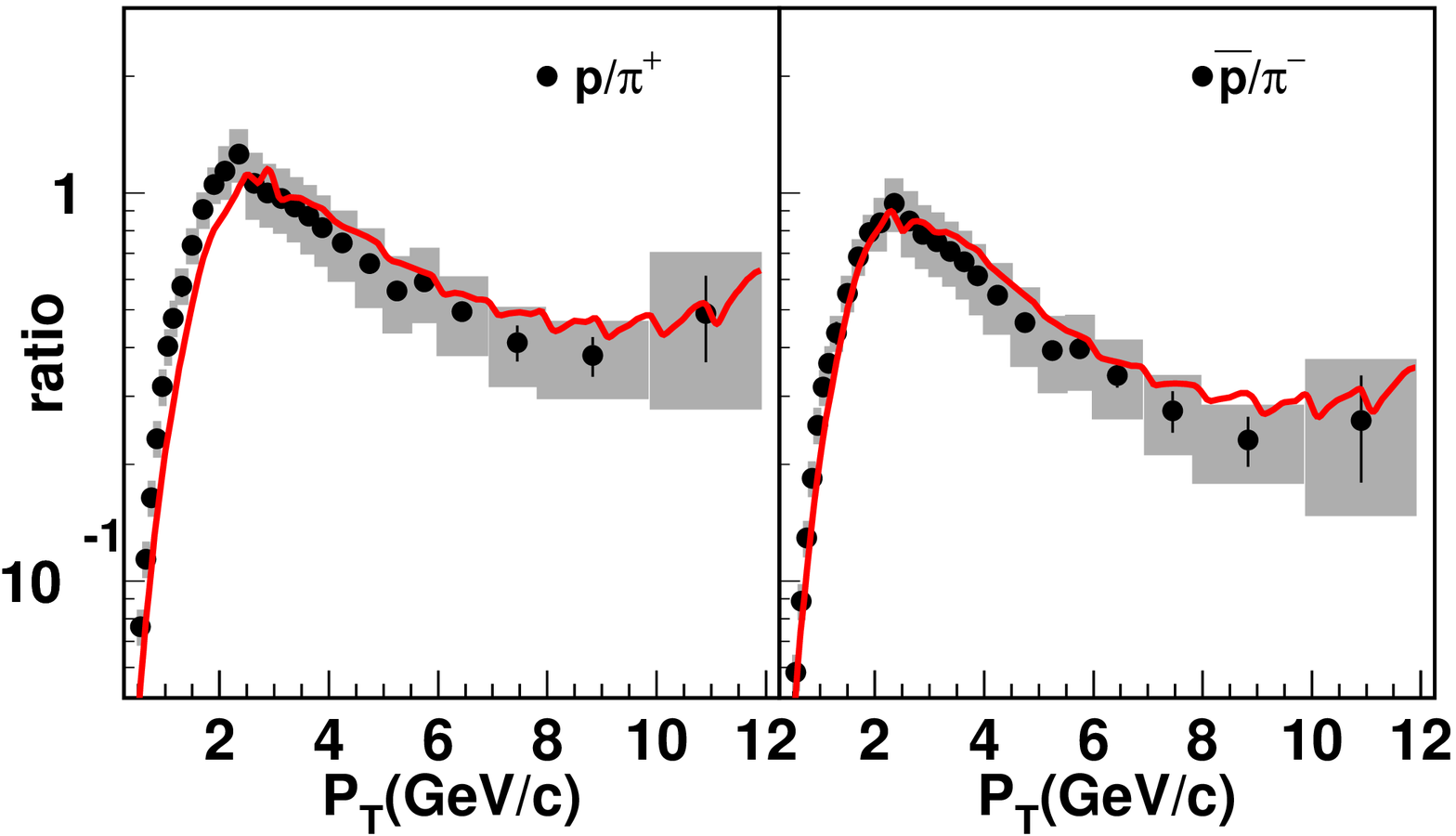,width=\linewidth} \caption{(Color online) The
ratios of $p/\pi^{+}$ and $\overline{p}/\pi^{-}$ as a function of
transverse momentum at midrapidity in central Au+Au collisions at
$\sqrt{s_{NN}}= 200 $ GeV. The shaded boxes represent the systematic
uncertainties. The solid lines are our results. Experimental data
are given by  STAR Collaboration  \cite{Abelev:2006}.} \label{ppi}
\end{figure}

Recently the STAR Collaboration has measured the $p/\pi^{+}$ and
$\overline{p}/\pi^{-}$ ratios at large transverse momenta in central Au+Au
collisions at $\sqrt{s_{NN}}= 200 $ GeV, and compared the data with the
predictions of recombination and coalescence model \cite{Abelev:2006}. It
is found that $p/\pi^{+}$ and $\overline{p}/\pi^{-}$ ratios peak at
$\mathrm {p_{T}}\sim 2-3$ GeV with values close to unity, decrease with
increasing $\mathrm {p_{T}}$, and approach
 $p/\pi^{+} \approx 0.4,\overline{p}/\pi^{-} \approx 0.25$ at
$\mathrm {p_{T}}\gtrsim 7$ GeV. These models can qualitatively describe the
$p(\overline{p})/\pi$ ratio at intermediate $\mathrm {p_{T}}$ but in general
underpredict the results at high $\mathrm {p_{T}}$ \cite{Abelev:2006}.
We also compute the $p/\pi^{+}$ and $\overline{p}/\pi^{-}$ ratios in the range
$\mathrm {p_{T}}<12$ GeV. The calculation results are shown in Fig.\ref{ppi}.
Our results agree with the experimental data in the full $\mathrm {p_{T}}$
range.

One can see from the above results that the quark combination model can reproduce
the productions of identified hadrons  quite well in a larger transverse momentum
range. It suggests that quark combination mechanism may still play an important role
at high transverse momentum up to 12 GeV/$c$.

\begin{figure*}[!tbp]
\epsfig{file=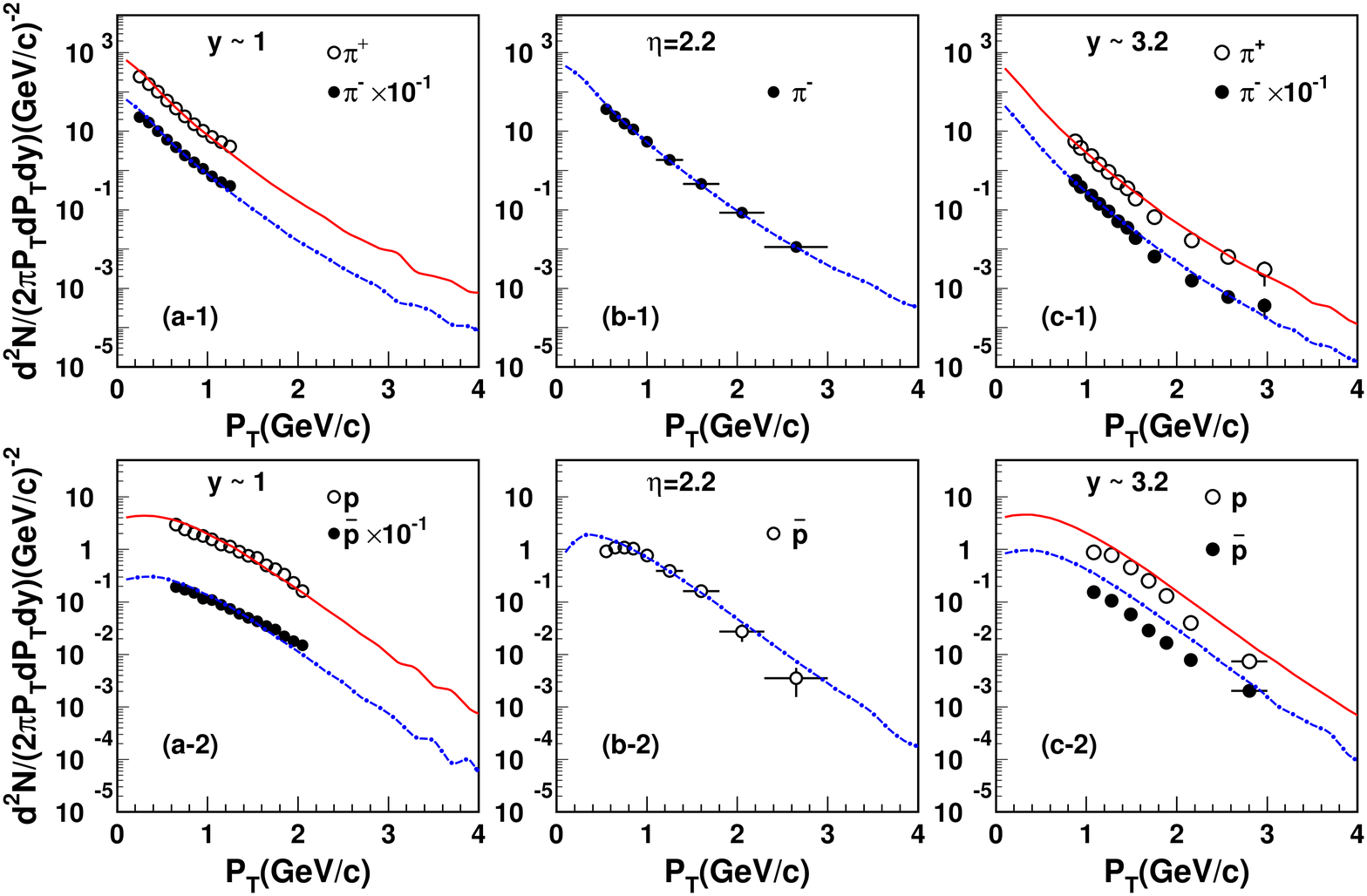,width=\linewidth} \caption{(Color online)
Transverse momentum spectra of pions and protons at $y\sim1$ (a),
$\eta=2.2$ (b) and $y=3.2$ (c) in central Au+Au collisions at
$\sqrt{s_{NN}}= 200 $ GeV. The solid lines are our results for
$\pi^{+}\text{and}\, p$, and the dotted-dashed lines are for
$\pi^{-}\text{and}\ \overline{p}$. The data of panel (a),(b),(c) are
given by  BRAHMS Collaboration
\cite{Arsene:2005,Arsene:2006,Radoslaw:2006}.} \label{ppion-y-1-2-3}
\end{figure*}
\subsection{Transverse momentum spectra of hadrons at other rapidities
$y\sim1$, $\eta=2.2$ and $y\thickapprox3.2$ }
A variety of experimental facts at RHIC indicate that the  QGP has
been probably produced at midrapidity. Based on this indication, quark combination picture
has successfully explained the hadron production at midrapidity
\cite{Hwa:2004,Greco:2003prc,Shao:2004cn}. However, can the QGP  extend
to forward rapidity? If it does, how much rapidity can it extend to?
Can combination picture describe the hadron production at forward
rapidity yet? Fortunately, the BRAHMS Collaboration recently has measured
the  $p_{T}$ spectra of identified hadrons at different rapidities
\cite{Arsene:2005,Arsene:2006,Radoslaw:2006}. It provides an opportunity
to study the hadrons production at different rapidities and rapidity dependence
of hadrons transverse momentum spectra. We have described the longitudinal
distributions of identified hadrons in section IV. The transverse momentum
spectra of hadrons at midrapidity are well reproduced in the above subsection.
We now  make a straightforward extension to the forward rapidity region,
and this extension should be made with no change in the quarks $\mathrm {p_{T}}$
spectra before we draw some useful physical information from it.

We compute the transverse momentum spectra of pions and protons at $y\sim1$
in central Au+Au collisions at $\sqrt{s_{NN}}= 200 $ GeV, the results are shown
in Fig\,\ref{ppion-y-1-2-3}(a). The proton and antiproton spectra are corrected
to remove the feed-down contributions from $\Lambda$ and $\overline{\Lambda}$
 weak decays. No significant changes for the transverse spectra of pions and
proton(antiproton) are observed within one unit around midrapidity. It suggests
that hadrons production at this rapidity are still dominated by combination
in intermediate transverse momentum range. The QGP still exist at $y\sim1$
and it may extend much more than this rapidity.

In Fig\,\ref{ppion-y-1-2-3}(b), we show the transverse momentum spectra of
pion and antiproton at $\eta=2.2\,(2.14<\eta<2.26)$ in central Au+Au collisions
at $\sqrt{s_{NN}}= 200 $ GeV. The correction for feed-down from the (anti-)lambda
decays has been applied, whereas the contamination of pions due to weak decays
has not been corrected. Our model can also well describe the hadronic
production at this rapidity.

We further compute the transverse spectra of  pions and (anti-)proton at
$y\thickapprox3.2$, the results are shown in Fig\,\ref{ppion-y-1-2-3}(c).
No correction for decay or feed-down has been applied to data. One can see
that computation results are in good agreement with the data for pions,  but
can't reproduce the data of proton and antiproton. This is probably because
the transverse momentum distribution of net-quarks varies with rapidity in
nucleus-nucleus collisions at RHIC.  The amount of net-quarks is marginal
compared with that of newborn quarks in midrapidity region. Net-quarks may be
highly thermalized  due to interactions with newborn quarks. Its transverse
momentum distribution will asymptotically  tend to that of newborn light quarks.
However the amount of net-quarks can approach and even exceed  that of newborn
quarks at large rapidity.  Therefore the thermalization extent of net-quarks at
large rapidity is much smaller than that at midrapidity. Its transverse momentum
distribution at large rapidity may be obviously different from that of newborn quarks.
Just as analyzed  in section IV, the yields of proton and antiproton are more
sensitive to the existence of net-quarks than that of pions. Thus the rapidity
dependence of transverse momentum distribution for net-quarks should be mainly
embodied by that of proton and antiproton, rather than pions. It is one of the
probable reason why our quark combination model can describe well the data of
pions  but for proton and antiproton at $y\thickapprox3.2$.

From the above results, we can see that the quark combination model is
able to describe the transverse momentum distribution of identified hadrons in
a broad rapidity range $0<y<3.2$. This implies that  quark combination
hadronization mechanism is still applicable to large rapidity (at least $y=3.2$).
On the other hand, the BRAHMS Collaboration has measured the nuclear
modification factor $R_{AA}$ for identified hadrons at forward
rapidity in central Au+Au collisions at $\sqrt{s_{NN}}= 200 $ GeV
\cite{Radoslaw:2006}, and the data also indicate the existence of
quark combination mechanism at forward rapidity. The continued
suppression seen in the $R_{AA}$ at $y=3.2$ further suggests that
there is different physics from that at midrapidity. The absence of
suppression in particle production at large transverse momentum in
d+Au collisions shows that the suppression in central Au+Au
collisions at midrapidity is not an initial-state effect but a
final-state effect of the produced density medium (jet quenching)
\cite{Adler:2003,Adams:2003}. However, the initial-state parton
saturation effects are more evident at large rapidity
\cite{Kharzeev:2003,Marian:2003}, which has been testified by the
high rapidity suppression measured in d+Au collisions
\cite{Arsene:2004}. But the decreasing of the parton medium density
at large rapidity will lead to the smaller jet energy loss, thus the
less suppression caused by jet quenching can be expected. Therefore
the suppression measured at large rapidity in central Au+Au
collisions may be the result of a compromise between initial- and
final-state effects \cite{Debbe:2006}. Maybe, the same parton
$\mathrm {p_{T}}$ spectra at $y=3.2$ with that at $y=0$ used in our
work is just a synthetic embodiment of the two effects.

\section{Summary}
Using the quark combination model, we study the rapidity and transverse
momentum distributions for identified hadrons in central Au+Au
collisions at $\sqrt{s_{NN}}= 200$ GeV, incorporating the collective
expansion of the hot medium. We use the measured rapidity spectra of $\pi^{+}$
and $K^{+}$ to extract the values of parameters and the explicit
form of the nonuniform expansion function. The rapidity
distributions of net-quarks are described by a relativistic
diffusion model \cite{wol99}. We compute the rapidity
distributions of identified hadrons. The results agree with the
experimental data. The existence of net quarks leads to the excess of
$K^{+}$ over $K^{-}$ and that of proton over antiproton in full
rapidity range. The difference between yields of $\pi^{+}$ and
$\pi^{-}$ resulting from net-quarks is very small compared to the
large yields of pions. We further predict the rapidity spectra of
$K^{0}_{s}$, $\Lambda(\overline{\Lambda} )$, $\mathrm{\Xi^{-}}$
($\mathrm{\overline{\Xi}^{\,_+}}$) and
$\mathrm{\Omega^{-}}+\mathrm{\overline{\Omega}}^{_+}$ in central
Au+Au collisions  at $\sqrt{s_{NN}}= 200$ GeV. We also calculate the
transverse spectra of identified hadrons and the ratios of proton/
antiproton to pions in a large transverse momentum range at
midrapidity. Our model reproduces the data quite well, which
suggests that combination scenario may still play an important role
in hadronization at high $p_{T}$ up to 12 GeV/$c$. Finally we give
rapidity dependence of transverse momentum spectra of hadrons. Using
the same transverse momentum spectra for quarks, we calculate
the transverse momentum spectra at various rapidities $y\sim0,1$,
$\eta=2.2$ and $y\thickapprox 3.2$. The results are in good
agreement with the data except those for proton and antiproton at
$y\thickapprox 3.2$.

\subsection*{ACKNOWLEDGMENTS}
The authors thank Q. Wang, S.-Y. Li, and Z.-T. Liang for helpful
discussions. The work is supported in part by the National Natural
Science Foundation of China under the grant 10475049, the foundation
of University Doctorate Educational Base of Ministry of Education
under the grant 20030422064, and the science fund of Qufu Normal
University.

\end{document}